\documentclass[11pt,a4paper,oneside]{article}
\usepackage[T1]{fontenc}
\usepackage[utf8]{inputenc}
\usepackage{lmodern}
\usepackage[a4paper,margin=25mm]{geometry}
\usepackage{microtype}
\usepackage{amsmath,amssymb}
\usepackage{graphicx}
\usepackage{xcolor}
\usepackage{authblk}
\usepackage{caption}
\usepackage{environ}
\usepackage{xparse}
\usepackage{ifthen}
\usepackage[numbers,sort&compress,round]{natbib}
\usepackage[hidelinks,breaklinks]{hyperref}
\date{}
\newcommand{\rev}[1]{#1}
\newcommand{\SI}[1]{\textit{#1}}
\makeatletter
\newcommand{\leadauthor}[1]{}
\newcommand{\significancestatement}[1]{\gdef\arxiv@significance{#1}}
\newcommand{\authorcontributions}[1]{\gdef\arxiv@contributions{#1}}
\newcommand{\authordeclaration}[1]{\gdef\arxiv@declaration{#1}}
\newcommand{\correspondingauthor}[1]{\gdef\arxiv@correspondence{#1}}
\newcommand{\keywords}[1]{\gdef\arxiv@keywords{#1}}
\let\arxiv@beginabstract\abstract
\let\arxiv@endabstract\endabstract
\RenewEnviron{abstract}{\global\let\arxiv@abstract\BODY}
\let\arxiv@originalmaketitle\maketitle
\renewcommand{\maketitle}{%
  \arxiv@originalmaketitle
  \begingroup\centering\footnotesize
    \arxiv@correspondence\par\smallskip
  \endgroup
  \begingroup\arxiv@beginabstract\arxiv@abstract\arxiv@endabstract\endgroup
  \begingroup\small\noindent\arxiv@keywords\par\endgroup
  \section*{Significance}
  \arxiv@significance\par\medskip
}
\newcommand{\matmethods}[1]{\gdef\arxiv@methods{#1}}
\newcommand{\showmatmethods}[1]{\section*{Materials and Methods}\arxiv@methods\par}
\newcommand{\acknow}[1]{\gdef\arxiv@acknowledgments{#1}}
\newcommand{\showacknow}[1]{%
  \section*{ACKNOWLEDGMENTS.}\arxiv@acknowledgments\par
  \section*{Author contributions:}\arxiv@contributions\par
  \section*{Competing interests}\arxiv@declaration\par
  \medskip\begingroup\small
  \noindent Copyright \textcopyright{} 2026 the Author(s). This open access article is distributed under Creative Commons Attribution-NonCommercial-NoDerivatives License 4.0 (CC BY-NC-ND).\par\smallskip
  \endgroup
}
\newboolean{shortarticle}
\newboolean{singlecolumn}
\setboolean{shortarticle}{false}
\setboolean{singlecolumn}{true}
\newcommand{\abscontentformatted}{}
\newcommand{\abscontent}{}
\let\ps@firststyle\ps@plain
\newcommand{\firstpage}[2][]{}
\newcommand{\dropcap}[1]{#1}
\newcommand{\bibsplit}[1][]{}
\makeatother
\newcommand{\sidecaptionrelwidth}{1}
\NewDocumentEnvironment{SCfigure*}{O{1} O{tbp}}
  {\begin{figure}[#2]}
  {\end{figure}}

\begin{document}

\title{Cell division sets a universal flow geometry in cell layers}

\author[a]{Tianxiang Ma}
\author[a]{Lasse Bonn}
\author[a]{Valeriia Grudtsyna}
\author[a]{Nigar Abbasova}
\author[a]{Martin Cramer Pedersen}
\author[b,c]{Nuno A. M. Araujo}
\author[a,1]{Amin Doostmohammadi}

\affil[a]{Niels Bohr Institute, University of Copenhagen, Copenhagen 2100, Denmark}
\affil[b]{Departamento de Física, Faculdade de Ciências, Universidade de Lisboa, Lisboa 1749-016, Portugal}
\affil[c]{Centro de Física Teórica e Computacional, Faculdade de Ciências, Universidade de Lisboa, Lisboa 1749-016, Portugal}

\leadauthor{Ma}

\significancestatement{Cell collectives can self-organize into large-scale flows whose vortical interfaces display universal symmetry across different living systems and appear similar under changes in position, orientation, and magnification. We show that cell division is a key process that maintains this organization. Beyond its role in growth, division preserves flexibility in packing and mechanical constraints. When division is blocked, large-scale motion persists, but the interface geometry no longer remains universal, indicating that these symmetry signatures require not only activity but also continual renewal of the cellular network. Our findings establish a direct link between a core biological process and physical symmetries, and suggest that antiproliferative interventions may disrupt symmetry-preserving states, altering collective mechanics and transport in epithelial and other systems.}

\authorcontributions{A.D. designed research; T.M., L.B., V.G., N.A., M.C.P., and N.A.M.A. performed research; T.M., L.B., V.G., N.A., and M.C.P. contributed new reagents/analytic tools; T.M., L.B., V.G., and N.A. analyzed data; and T.M. and A.D. wrote the paper.}
\authordeclaration{The authors declare no competing interest.}
\correspondingauthor{\textsuperscript{1}To whom correspondence may be addressed. Email: \href{mailto:doostmohammadi@nbi.ku.dk}{doostmohammadi@nbi.ku.dk}.}


\begin{abstract}
Collective flows in epithelial tissues contain a geometric backbone of vortical interfaces whose statistics exhibit hallmarks of critical percolation and conformal invariance. Yet how fundamental cellular processes govern the breakdown of such symmetry-rich flow geometry remains unclear. Here we show that cell division plays a central physical role in regulating this universal flow organization by controlling both the geometric and mechanical flexibility of the cell–cell network. Using pharmacological perturbations, we find that when proliferation is suppressed through two independent interventions, coherent flows persist but neighbor exchanges decline and conformally invariant geometry is lost. 
Blocking apoptosis does not affect universality, isolating division as the key control. A vertex model with tunable division quantitatively reproduces these effects and restores conformal invariance when division is allowed. 
We further trace this effect to changes in both the geometric and mechanical organization of the cell layer: Divisions act as intermittent topological renewals that loosen constraints, preserving the network’s flexibility and capacity to rearrange across scales. 
Thus, beyond its canonical role in growth, cell division acts as a structural and mechanical regulator of collective self-organization. These findings establish a direct connection between fundamental biological processes and emergent physical symmetries.
\end{abstract}

\maketitle
\thispagestyle{firststyle}
\ifthenelse{\boolean{shortarticle}}{\ifthenelse{\boolean{singlecolumn}}{\abscontentformatted}{\abscontent}}{}

\firstpage[1]{4}

\section*{Introduction}
\dropcap{C}ollective flows in epithelial monolayers underlie essential biological processes such as tissue development, wound healing, and cancer invasion~\cite{friedl2009collective,angelini2011glass,tetley2019tissue,palamidessi2019unjamming,ilina2020cell,bruckner2024learning,cheung2025collective}. These flows arise from the interplay between active forces generated by individual cells and cell--cell interactions, mediated through mechanical and biochemical interactions~\cite{thuroff2019bridging,alert2020physical,balasubramaniam2021investigating,ziepke2022multi,gompper20252025,volpe2025roadmap}. Such collective dynamics are now central to the study of active matter~\cite{wensink2012meso,marchetti2013hydrodynamics,goldstein2015green,doostmohammadi2018active,fodor2022irreversibility,gompper20252025}, offering a biologically grounded platform to investigate principles of nonequilibrium statistical mechanics.

A central theme emerging across living systems is that their collective behavior can become independent of microscopic details, following statistical laws shared with equilibrium critical phenomena~\cite{stanley1996scaling,mora2011biological,petridou2021rigidity,sun2025feedback}.
Such scale-invariant statistics link biological function to collective organization rather than to the properties of individual cells~\cite{bogdan2021biological,sole2024fundamental}.
Similar signatures appear in collective flows~\cite{doostmohammadi2017onset,wei2024scaling,perez2025bacteria,andersen2025evidence}, where large-scale patterns exhibit conserved statistical structure that transcends cellular specifics.
Recently it was found that across diverse systems, including epithelial monolayers and bacterial collectives, the zero-vorticity interfaces, that form the geometric backbone of the flow field, obey the strong symmetry constraints of conformal invariance, encompassing translational, rotational, and scale symmetries, consistent with the universality class of two-dimensional critical percolation~\cite{andersen2025evidence}.
This means that, even within intrinsically far-from-equilibrium flows of living systems, a geometric backbone emerges that obeys the same universal statistics as equilibrium critical systems.

The emergence of such equilibrium-like features in an active, far-from-equilibrium tissue flow raises a conceptual puzzle: Under what perturbations of fundamental cellular processes do these signatures break down? And what mechanisms underlie this breakdown?
In living tissues, the continuous renewal of cells through division and elimination is a fundamental process that regulates collective organization and, consequently, collective flow~\cite{rossen2014long,czajkowski2019glassy,li2021role,devany2021cell,bocanegra2023cell}.
More importantly, the processes of division and elimination are conserved across diverse biological systems and, although governed by cell-specific mechanisms at the microscopic level, they give rise to emergent dynamics at the tissue scale that are insensitive to microscopic details~\cite{gelimson2015collective,klingel2025population}. 
We therefore focus on cell division and elimination as intrinsic biological processes and examine their role in maintaining the universal statistical structure of collective flows. 

Using pharmacological suppression of cell division and apoptosis, we demonstrate that division, rather than elimination, is essential for preserving conformally invariant organization in epithelial flows. 
Combining division-controlled simulations with geometric and mechanical analyses, we identify a mechanism by which cell division regulates the flow backbone by maintaining a tissue architecture that supports fluid-like rearrangements.

\section*{Results}
\subsection*{Inhibition of Cell Division Disrupts Conformal Invariance and Percolation Universality.}
We asked whether the geometric backbone of epithelial flows, previously associated with the critical percolation universality class~\cite{andersen2025evidence}, is maintained when cell division is suppressed.
To test this, we treated confluent MDCK monolayers with Mitomycin C (see details in \SI{SI Appendix, Methods}), which inhibits DNA synthesis and arrests the cell cycle~\cite{siedlik2017cell,matsiaka2018discrete,gauquelin2019influence}.
Division suppression was confirmed by stable cell density, whereas control monolayers show a gradual increase in density over time (\SI{SI Appendix, Fig.~S1A}), and by significantly reduced division rates (\SI{SI Appendix, Fig.~S1B}).
We also verified that Mitomycin C does not substantially alter other turnover pathways, as cell extrusion rates remain comparable to control (\SI{SI Appendix, Fig.~S1C}).
Collective flows were quantified from time-lapse phase-contrast microscopy using Particle Image Velocimetry (PIV) with identical image-processing parameters applied to both control and Mitomycin C–treated monolayers \cite{thielickePIVlabUserfriendlyAffordable2014}.

\begin{SCfigure*}[\sidecaptionrelwidth][t!]
\centering
\includegraphics[width=11.4cm]{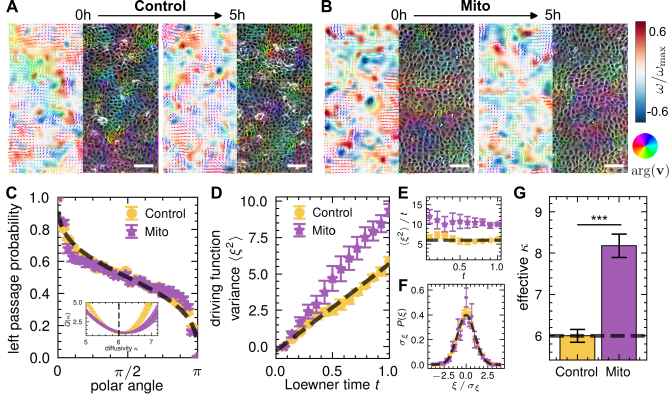}
\caption{
\textbf{Conformal invariance and universality are altered by blocking cell division.}
\textbf{(A and B)} Representative control \textbf{(A)} and Mitomycin C (Mito)-treated \textbf{(B)} MDCK monolayers at 0 and 5 h, overlaid with velocity vectors (colored by orientation) and vorticity fields. Vorticity is shown as normalized $\omega/\omega_{\max}$. (Scale bars, 100~$\mu$m.)
\textbf{(C)} Left-passage probability versus polar angle. Control closely follows the SLE$_6$ prediction with diffusivity $\kappa = 6$ (dashed black line), whereas Mito deviates. Inset: weighted mean-square deviation $Q(\kappa)$, with a minimum near $\kappa = 6$ in control.
\textbf{(D)} Driving function variance versus Loewner time. The dashed black line marks the SLE$_6$ reference slope ($\kappa=6$, for critical percolation). Control is consistent with $\kappa=6$ ($P=0.803$), whereas Mito deviates significantly ($P=1.29\times10^{-3}$).
\textbf{(E)} Driving function variance normalized by Loewner time \cite{bernard2007inverse}, showing that the effective diffusivity remains close to 6 in control but deviates in Mito.
\textbf{(F)} Probability density of the driving function $\xi$, rescaled by its SD $\sigma_{\xi}$.
\textbf{(G)} Effective $\kappa$, obtained for each replicate by averaging the estimates from $Q(\kappa)$ and the driving function variance slope. Mito is significantly higher than control (Welch’s two-sample $t$ test, ***$P<0.001$).
All data represent mean ± SE of the mean (SEM) from 5 independent experiments.
}
\label{fig:fig1}
\end{SCfigure*}

Representative velocity and vorticity fields illustrate that large-scale flow structures persist even when cell division is blocked. \hyperref[fig:fig1]{Fig.~1~A and~B} shows bright-field images overlaid with velocity and vorticity fields: Both untreated (Control) and Mitomycin C-treated (Mito) monolayers exhibit jets and vortices, indicating that collective motion continues despite division suppression (\SI{Movies~S1~and~S2}). Cell proliferation therefore is not required to generate large-scale collective flows.

The mere presence of coherent flows does not imply that their underlying geometry preserves conformal symmetry.
To test this, we focused on the zero-vorticity isolines (also known as nodal lines of the vorticity field) that separate clockwise and counterclockwise rotating regions (\SI{SI Appendix, Fig.~S2}). These contours form a geometric backbone of the flow, whose statistical properties can be compared against known universality classes from two-dimensional critical phenomena~\cite{andersen2025evidence}. Specifically, we analyzed whether the vorticity contours display conformal invariance, using the framework of Schramm–Loewner Evolution (SLE). 
SLE provides a stringent test of conformal invariance by mapping interface geometry onto a stochastic process characterized by a single parameter $\kappa$. SLE has been rigorously proven for a number of critical systems, including critical percolation~\cite{smirnov2001critical}, the Q-state Potts model~\cite{Rohde05}, loop-erased random walks, and the uniform spanning tree~\cite{lawler2005conformally}, and strong numerical and experimental evidence supports SLE behavior in systems such as turbulent fluid interfaces~\cite{bernard2006conformal}, water wave-turbulence~\cite{noseda2024conformal}, watersheds~\cite{daryaei2012}, liquid crystals~\cite{almeida2025critical}, and rigidity percolation~\cite{Javerzat2024}.

The results of this analysis are summarized in \hyperref[fig:fig1]{Fig.~1~C--G}. 
In control monolayers, the vorticity contours display the defining signatures of conformal symmetry.
The left-passage probability, which measures the probability that a contour passes to the left of a given spatial point, matches the theoretical prediction for SLE$_6$, which corresponds to critical percolation (\hyperref[fig:fig1]{Fig.~1C}). Correspondingly, the weighted mean-square deviation $Q(\kappa)$, which quantifies the deviation between the measured and theoretical left-passage probabilities for different $\kappa$ values, exhibits a sharp minimum at $\kappa \approx 6$ (Inset of \hyperref[fig:fig1]{Fig.~1C}).
In addition, we performed an independent consistency check using the driving function from the Loewner evolution, which represents the stochastic process generating the contour under conformal mapping. In the control condition, the driving function variance grows linearly with Loewner time with a slope near 6 and is statistically consistent with the SLE$_6$ reference value (slope $=5.924 \pm 0.285$; $P=0.803$; \hyperref[fig:fig1]{Fig.~1D}), providing independent support for SLE$_6$ signatures.

By contrast, Mitomycin C-treated monolayers show stronger deviations from these features. The left-passage probability deviates from the SLE$_6$ prediction (\hyperref[fig:fig1]{Fig.~1C}). The weighted mean-square deviation $Q(\kappa)$ deviates from the minimum at $\kappa = 6$ and instead shifts toward higher values (Inset of \hyperref[fig:fig1]{Fig.~1C}).
Furthermore, the variance of the driving function deviates from the SLE$_6$ expectation: The linear fit yields a slope significantly different from 6 (slope $=10.118 \pm 0.511$; $P=1.29\times10^{-3}$; \hyperref[fig:fig1]{Fig.~1D}), indicating a significant departure from SLE$_6$. Consistently, when the variance is normalized by Loewner time \cite{bernard2007inverse, bernard2006conformal}, Mitomycin C-treated monolayers no longer remain near the theoretical level of 6, unlike the control condition (\hyperref[fig:fig1]{Fig.~1E}). In addition, the driving function loses Gaussianity under Mitomycin C treatment (\hyperref[fig:fig1]{Fig.~1F}), further supporting its departure from the SLE$_6$ universality class.

\begin{SCfigure*}[\sidecaptionrelwidth][t!]
\centering
\includegraphics[width=11.4cm]{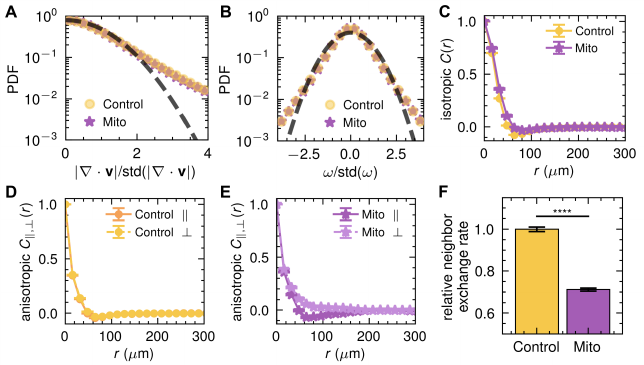}
\caption{
\textbf{Cell division sustains isotropic flow organization and promotes tissue fluidity.}  
\textbf{(A and B)} Probability density distributions of normalized absolute divergence \textbf{(A)} and vorticity \textbf{(B)} for control and Mitomycin C (Mito)-treated monolayers, compared to a standard normal distribution (dashed black line).  
\textbf{(C)} Vorticity autocorrelation functions $C(r)$ exhibit similar spatial decay in both conditions (mean ± SEM, 5 independent experiments).
\textbf{(D and E)} Anisotropic vorticity correlations parallel ($\parallel$) and perpendicular ($\perp$) to local velocity directions for \textbf{(D)} control and \textbf{(E)} Mito-treated monolayers. Control cells show isotropic decay, while Mito-treated cells display anisotropy with longer-range perpendicular correlations. 
\textbf{(F)} Neighbor exchange rates, defined as the number of exchanges per cell per unit time normalized by total neighbor count. Control monolayers exhibit significantly higher rates than Mito-treated monolayers. Bar plots show values expressed relative to control (Welch’s two-sample $t$ test; ****$P < 0.0001$). 
All data represent mean ± SE of the mean (SEM) from 5 independent experiments.
}
\label{fig:fig2}
\end{SCfigure*}

Finally, we estimated the effective diffusivity $\kappa$ for each replicate by averaging the values obtained from the minimum of $Q(\kappa)$ in the left-passage analysis and from the slope of the driving function variance analysis. As shown in \hyperref[fig:fig1]{Fig.~1G}, the combined estimate shows that the control condition is statistically consistent with the SLE$_6$ prediction ($6.002 \pm 0.155$, $P=0.9903$), whereas Mitomycin C treatment yields a significantly larger effective diffusivity ($8.182 \pm 0.285$), differing significantly from both the theoretical value ($P=1.56\times10^{-3}$) and the control condition ($P=4.7\times10^{-4}$).

To further verify this deviation, we tested scale invariance, the necessary prerequisite for conformal invariance. The vorticity-cluster size distributions enclosed by zero-vorticity isolines show that the control condition is consistent with the expected scale-invariant power law and critical-percolation exponent, whereas Mitomycin C-treated monolayers deviate significantly (\SI{SI Appendix, Fig.~S3A}). Likewise, the generalized fractal dimensions remain close to the SLE$_6$ value of $7/4$ and nearly independent of moment order in control, but start around 1.58 and vary with moment order under Mitomycin C treatment, indicating multifractality and loss of scale-invariant geometry \cite{bernard2007inverse} (\SI{SI Appendix, Fig.~S3B}).

To test the robustness of these conclusions on the role of cell division, we treated cell monolayers with a different pharmacological agent, Nocodazole, which depolymerizes microtubules and prevents spindle formation \cite{jordan1998microtubules, cho2006analysis} (see details in \SI{SI Appendix, Methods}).
Nocodazole-treated monolayers show similar deviations: The left-passage probability deviates from the SLE$_6$ prediction (\SI{SI Appendix, Fig.~S4~A and~B}), and the driving function shows nonlinear variance growth (\SI{SI Appendix, Fig.~S4C}). Together, these results demonstrate that, while large-scale collective motion persists, conformal symmetry and the universal flow geometry of the vorticity backbone are lost when cell division is inhibited.

\subsection*{Apoptosis Does Not Affect Universal Flow Geometry.}
Because both cell division and apoptosis contribute to tissue renewal and can influence collective organization, we next asked whether apoptosis plays a comparable role in maintaining universal flow geometry. Beyond its biological relevance, this test was essential to validate the interpretation of the division-inhibition experiments, since both Mitomycin C and Nocodazole can influence not only proliferation but also apoptotic activity through their coupling with the cell cycle, density regulation, and microtubule dynamics~\cite{pirnia2002mitomycin,beswick2006nocodazole}.
To this end, we treated monolayers with the apoptosis inhibitor Q-VD-OPh (see details in \SI{SI Appendix, Methods}), which suppresses apoptosis while leaving cell division rates close to control (\SI{SI Appendix, Fig.~S1~B and~C}). Remarkably, under this condition, all hallmark signatures of conformal symmetry were preserved (\SI{SI Appendix, Fig.~S5}), including the left-passage probability and the linear variance growth of the driving function, both consistent with SLE$_6$. These results demonstrate that the loss of conformal symmetry arises specifically from suppression of cell division rather than from apoptosis.

\subsection*{Vortical Flows Become Anisotropic upon Mitomycin C Treatment but Not with Nocodazole.}

To investigate the mechanism by which cell division inhibition disrupts the universal features of flow backbone, we first examined structural and dynamical features of cellular flows. A key observation is that the overall degree of incompressibility---quantified by the absolute divergence of the velocity field---remains comparable between control and division-inhibited monolayers (Mitomycin C-treated in \hyperref[fig:fig2]{Fig.~2A}; Nocodazole-treated in \SI{SI Appendix, Fig.~S6A}). Similarly, the statistical distribution of vorticity is unchanged across conditions (Mitomycin C-treated in \hyperref[fig:fig2]{Fig.~2B}; Nocodazole-treated in \SI{SI Appendix, Fig.~S6B}) and remains strongly non-Gaussian, indicating that coherent vortical structures and active flow intermittency persist even when cell division is suppressed. In addition, the spatial vorticity–vorticity correlation functions $C(r)$ are nearly identical (Mitomycin C-treated in \hyperref[fig:fig2]{Fig.~2C}; Nocodazole-treated in \SI{SI Appendix, Fig.~S6C}). These controls indicate that division suppression does not alter the gross statistics of flow compressibility or vorticity.

After confirming that the overall statistics of cellular flow do not change, the most direct test for understanding the loss of conformal invariance is to examine whether the symmetry of the flow field itself is affected by division inhibition, since conformal invariance fundamentally relies on strict symmetry constraints, which includes translational, rotational, and scale invariance.
We find that anisotropic flow structures emerge in Mitomycin C–treated monolayers.
Specifically, we computed vorticity–vorticity correlation functions separately along directions parallel and perpendicular to the local velocity vector. In control monolayers, correlations are isotropic: Parallel and perpendicular directions exhibit indistinguishable decay profiles (\hyperref[fig:fig2]{Fig.~2D}). In contrast, Mitomycin C-treated monolayers display significantly longer correlation lengths in the direction perpendicular to the local flow (\hyperref[fig:fig2]{Fig.~2E}), indicating a breakdown of rotational symmetry and the emergence of flow anisotropy. Because conformal invariance encompasses rotational symmetry, this loss of rotational symmetry provides one route by which universality can be destabilized.

To test whether this anisotropic flow pattern represents a general consequence of division inhibition or a treatment-specific effect, we examined Nocodazole-treated monolayers. 
Surprisingly, no such anisotropy was detected in this condition (\SI{SI Appendix, Fig.~S7A}), suggesting that while anisotropic correlations can explain the breakdown of conformal invariance in Mitomycin C–treated monolayers, an alternative and more fundamental mechanism likely underlies the loss of conformal symmetry and universality. 

\begin{SCfigure*}[\sidecaptionrelwidth][t!]
\centering
\includegraphics[width=11.4cm]{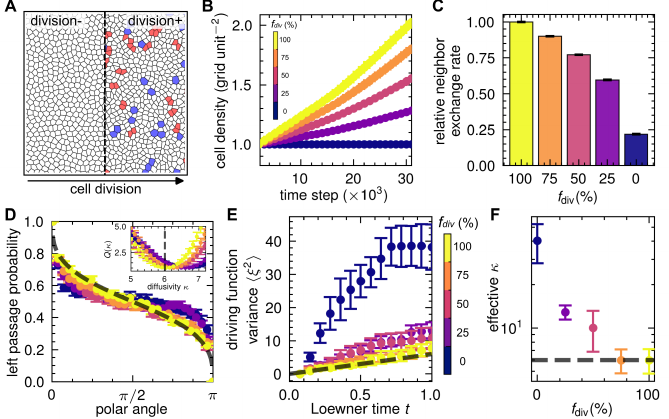}
\caption{
\textbf{Cell division sustains neighbor exchange and conformal invariance in vertex model simulations.}  
\textbf{(A)} Snapshots from vertex model simulations comparing monolayers without and with division, where red marks newly born cells and blue marks cells undergoing division in this frame.
\textbf{(B)} Time evolution of cell density for division strengths $f_\mathrm{div}$ ranging from 0\% (blue) to 100\% (yellow). 
\textbf{(C)} Relative neighbor exchange rates, normalized to the 100\% division case, decrease with lower $f_\mathrm{div}$. 
\textbf{(D)} Left-passage probability as a function of polar angle matches the SLE$_6$ prediction at $f_\mathrm{div}=100\%$ and progressively deviates as $f_\mathrm{div}$ decreases. The dashed black line shows Cardy’s formula for critical percolation. Inset: weighted mean-square deviation $Q(\kappa)$ curves confirm the trend, exhibiting a sharp minimum near $\kappa = 6$ at high $f_\mathrm{div}$ and shifting away from this value as $f_\mathrm{div}$ decreases.  
\textbf{(E)} Driving function variance grows linearly with Loewner time with a slope consistent with $\kappa=6$ (dashed line) at $f_\mathrm{div}=100\%$, but deviates at lower $f_\mathrm{div}$.
\textbf{(F)} Effective diffusivity $\kappa$ as a function of division strength $f_\mathrm{div}$.
All data represent mean ± SE of the mean (SEM) from 5 independent simulations.
}
\label{fig:fig3}
\end{SCfigure*}

\subsection*{Reduced Rearrangement Capacity Underlies the Breakdown of Conformal Symmetry.}

The absence of anisotropy in Nocodazole-treated tissues suggested that a more general mechanism must account for the loss of conformal symmetry and universality.
We therefore turned to the organization of the cell–cell contact network, whose ability to rearrange determines whether the tissue can explore different topological configurations. 
To test this, we quantified neighbor exchange rates as a direct measure of the rearrangement capacity of the cell layer. 
Control monolayers exhibited frequent neighbor exchanges, whereas Mitomycin C-treated monolayers showed a marked reduction (\hyperref[fig:fig2]{Fig.~2F}). A similar suppression was observed in Nocodazole-treated monolayers (\SI{SI Appendix, Fig.~S7B}), indicating that while the collective flow remains, the cellular network becomes more constrained. 
In addition, we quantified structural relaxation by computing the self-intermediate scattering function and extracting the $\alpha$-relaxation time $\tau_\alpha$~\cite{das2021controlled}. Consistent with the reduced neighbor exchange rate, $\tau_\alpha$ increases upon Mitomycin C and Nocodazole treatment (\SI{SI Appendix, Figs.~S8 and~S9}), suggesting slower structural relaxation of the monolayer under division inhibition.
Thus, the breakdown of conformal invariance is not due to a loss of motion or a transition to a solid-like jammed state, as the system remains dynamically active and supports vortical flows, but rather to the reduction in continuous rearrangements within the cell layer.

To further confirm this hypothesis, we turned to physical modeling using a two-dimensional vertex model of epithelial monolayers. In this framework, cells are represented as polygons, and neighbor exchanges occur through T1 transitions, where a shared edge between adjacent cells shortens, collapses, and reforms in a new orientation.
In our implementation, stochastic line-tension fluctuations facilitate T1 events~\cite{krajnc2020solid, bocanegra2023cell,arzash2024tuning}. We chose the fluctuation strength such that the mean shape index lies above the solid–fluid threshold $p^*=3.81$~\cite{bi2015density,krajnc2020solid} (\SI{SI Appendix, Fig.~S10}).
Building on recent work by Guerrero \textit{et al.}~\cite{guerrero2019neuronal} and Bocanegra-Moreno \textit{et al.}~\cite{bocanegra2023cell}, we incorporated cell division by tracking cell lineages and assigning each cell a time-dependent target area that evolves through the cell cycle. A cell divides once its target area exceeds a threshold in the mitotic phase~\cite{sorce2015mitotic, nematbakhsh2017multi}.

To enable tunable control over division frequency, we introduced a probabilistic division parameter $f_\mathrm{div}$, such that only a fraction of cells enter the cell cycle and divide upon reaching mitosis, while others retain their size and lineage. Simulation snapshots at different division probabilities are shown in \hyperref[fig:fig3]{Fig.~3A}, with the corresponding evolution of cell density over time in \hyperref[fig:fig3]{Fig.~3B} (see also simulation movies of different $f_\mathrm{div}$ in \SI{Movie~S3}).  
Simulations were performed under periodic boundary conditions; however, boundary regions were cropped during analysis to avoid edge artifacts and to match the nonperiodic configuration of experimental imaging. 
This formulation allowed us to vary division rate while holding all other tissue parameters constant (full details in \SI{SI Appendix, Methods}).

Even without division, the simulated monolayers remain motile and display vortical flows, confirming that activity and motion persist.
However, the frequency of neighbor exchanges declines as the division parameter $f_\mathrm{div}$ decreases (\hyperref[fig:fig3]{Fig.~3C}), mirroring the experimental reduction in network rearrangement shown in \hyperref[fig:fig2]{Fig.~2F}. Consistent with this trend, structural relaxation becomes faster at higher $f_\mathrm{div}$, with a systematic decrease in the $\alpha$-relaxation time $\tau_\alpha$ (\SI{SI Appendix, Fig.~S11}).

We further assessed conformal invariance by extracting the simulated vorticity fields (see vorticity evolution movies for different $f_\mathrm{div}$ in \SI{Movie~S4}) and applying the same SLE-based analysis used for the experiments. The results are summarized in \hyperref[fig:fig3]{Fig.~3~D--F}. At $f_\mathrm{div} = 100\%$, the simulated monolayers recapitulate key features of universal behavior and conformal invariance: The left-passage probability as a function of angle matched the theoretical SLE$_6$ prediction (\hyperref[fig:fig3]{Fig.~3D}), the weighted mean-square deviation $Q(\kappa)$ exhibited a sharp minimum near $\kappa = 6$ (Inset of \hyperref[fig:fig3]{Fig.~3D}), and the variance of the driving function grows linearly with Loewner time with a slope consistent with $\kappa=6$ (\hyperref[fig:fig3]{Fig.~3E}). Both diagnostics show progressive deviations as division strength decreases. 
Accordingly, as shown in \hyperref[fig:fig3]{Fig.~3F}, the effective diffusivity $\kappa$ exhibits a gradual deviation at intermediate division strengths ($f_{\mathrm{div}}\approx 50\%$) and a pronounced breakdown at lower $f_{\mathrm{div}}$, consistent with the experimentally observed loss of conformal symmetry upon division inhibition (\hyperref[fig:fig1]{Fig.~1~C--G}). Notably, increasing the target shape index, and thereby promoting baseline tissue rearrangements, shifts the onset of SLE$_6$-like signatures to lower $f_\mathrm{div}$ (\SI{SI Appendix, Fig.~S12}).

To test robustness, we implemented an alternative scheme for varying cell division in which cell target area was independent of the cell cycle and instead fluctuated randomly~\cite{bocanegra2023cell}. When a cell entered mitosis and its area exceeded the defined threshold, it underwent division (\SI{SI Appendix, Fig.~S13~A and~B}). Consistent with the main model, cell collectives with division exhibited significantly higher neighbor exchange rates than those without (\SI{SI Appendix, Fig.~S13C}). Consistent with this, SLE analyses revealed that suppression of cell division altered both the left-passage probability (\SI{SI Appendix, Fig.~S13D}) and the variance of the driving function (\SI{SI Appendix, Fig.~S13E}). 
Finally, mirroring the experimental approach, we suppressed cell division ($f_\mathrm{div}=0\%$) and varied the strength of cell removal (see \SI{SI Appendix, Methods} for details). These analyses showed that, unlike cell division, cell removal alone could not restore SLE$_6$-like behavior (\SI{SI Appendix, Fig.~S14}).

Taken together, these results demonstrate that intermittent divisions maintain a rearrangement capacity that is necessary for the flow field to sustain the conformally invariant flow geometry. This conclusion holds across both experimental methods of division inhibition and distinct implementations of division in vertex model simulations.

\subsection*{Geometric and Mechanical Responses Link Cell Division to Conformal Symmetry of the Collective Flow.}

Having established that cell division reduces neighbor exchange and disrupts conformal symmetry, we next asked what mechanism gives rise to this change. Because the ability of a tissue to undergo rearrangements depends on how cells are arranged, we hypothesized that division alters both the intercellular geometric and mechanical states, thereby translating into changes in flow organization.

\begin{figure}%[tbhp]
    \centering
    \includegraphics[width=11.4cm]{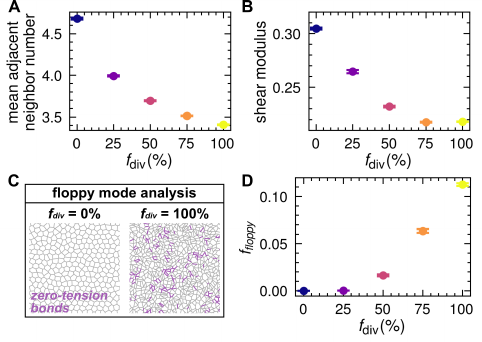}
    \caption{\textbf{Cell division jointly tunes geometry and mechanics.}
    \textbf{(A)} Mean adjacent neighbor number (mean degree of the cell-center adjacency graph) decreases with $f_\mathrm{div}$, showing a sharp drop up to $f_\mathrm{div}=50\%$ followed by a plateau.
    \textbf{(B)} Direct mechanical measurements show progressive softening: The instantaneous shear modulus decreases with $f_\mathrm{div}$, with a matching crossover near $f_\mathrm{div}\!\approx\!50\%$.
    \rev{\textbf{(C)} Representative snapshots of zero-tension bonds in the vertex model at different division strengths. Purple lines indicate bonds with zero tension, while gray lines show all bonds.}
    \textbf{(D)} The fraction of floppy modes increases with $f_\mathrm{div}$.
    All data represent mean ± SE of the mean (SEM) from 5 independent simulations.
    }
    \label{fig:fig4}
\end{figure}

To test this, we first quantified the intercellular geometric packing by constructing the cell-center adjacency graph from simulated cell centroids across varying values of the division parameter $f_\mathrm{div}$ (see details in \SI{SI Appendix, Methods}). 
We found that the \rev{mean adjacent neighbor number} decreases sharply as $f_\mathrm{div}$ increases from 0 to 50\%, after which it plateaus (\hyperref[fig:fig4]{Fig.~4A} and \SI{SI Appendix, Fig.~S15A}). In addition, the fraction of cells belonging to the \rev{largest adjacent component} shows the same trend (\SI{SI Appendix, Figs.~S15 and~S16}). These results show that geometry of the cellular network is altered once division changes, the same threshold coincides with the SLE analysis, in which conformal features largely deviated below $f_\mathrm{div} \approx 50\%$.
Consistent with the simulations, experimental data from Mitomycin C–treated and Nocodazole-treated monolayers show the same trend: Inhibited division increases \rev{mean adjacent neighbor number} and raises the fraction of cells belonging to the \rev{largest adjacent component} (\SI{SI Appendix, Fig.~S17}).

Motivated by prior work showing that packing and topology are tightly linked to the mechanics in epithelia~\cite{popovic2021inferring,haas2025geometry}, we next probed the mechanical state of the vertex-model tissue directly~\cite{staple2010mechanics, yan2019multicellular, duclut2021nonlinear,staddon2023role,damavandi2025universality}. 
We found both the instantaneous shear modulus and Young’s modulus decrease systematically with increasing $f_{\mathrm{div}}$ (\hyperref[fig:fig4]{Fig.~4B} and \SI{SI Appendix, Fig.~S18A}), indicating progressive softening. Consistent with this, the fraction of floppy modes increases with $f_{\mathrm{div}}$ (\hyperref[fig:fig4]{Fig.~4~C and~D}), indicating an increasing abundance of soft (low-energy) deformation modes and the signature of reduced rigidity~\cite{yan2019multicellular}. 
Together, these measurements show that division renders the tissue mechanically softer and more easily deformed on short timescales. Under continuous driving (line-tension fluctuations and ongoing divisions), such softening facilitates cell motions and can promote overall rearrangement.
In addition to these instantaneous responses, we quantified the T1 transition energy barrier~\cite{bi2014energy,bi2015density} and found that it decreases with $f_{\mathrm{div}}$ (\SI{SI Appendix, Fig.~S18B}), indicating that division lowers the energetic threshold for the elementary rearrangements that underpin tissue remodeling over longer timescales~\cite{krajnc2018fluidization}.
Notably, all of these mechanical readouts exhibit a robust crossover near $f_{\mathrm{div}}\approx50\%$, consistent with the threshold identified by the geometric and SLE analyses.

These results confirm that suppressing cell division constrains both the geometry and the mechanics of the cell layer. Although collective motion persists when division is inhibited, the ability of the cell--cell network to explore configurations is reduced. The system therefore remains dynamically active but overconstrained, breaking the rotational and scale symmetries required for conformal invariance.

\section*{Discussion}
Our findings reveal that while collective flows appear preserved in epithelial monolayers with inhibited proliferation, 
the signatures of conformal invariance and critical percolation statistics of the nodal lines of the vorticity field are lost.
Notably, this shift is not due to mere density changes or passive tissue renewal: When apoptosis was pharmacologically inhibited, conformal symmetry was preserved, even though overall density evolution was also altered.
These findings establish that the geometric order of epithelial flows is an emergent consequence of regulated cellular processes, specifically division-driven tissue renewal, and not a passive outcome of collective movement.
By combining two different pharmacological perturbations with a vertex model in which division frequency could be tuned in two independent ways, we identified a single mechanistic chain linking geometric packing, mechanical response, cell rearrangements, and the resulting shifts in the symmetry diagnostics of the flow backbone.
In this picture, cell division modulates intercellular architecture, preventing the tissue from becoming overconstrained and thereby maintaining the structural and mechanical flexibility required for rearrangements to propagate across the tissue.

From a physics standpoint, these results reveal how a far-from-equilibrium system can sustain universal and conformally symmetric flow organization through the organization of its microscopic constituents. \rev{A widely recognized physical role of cell division in collective flow is to enhance fluidity \cite{ranft2010fluidization,petridou2019fluidization,bocanegra2023cell};} however, the deviation of the conformal signatures cannot be attributed to the fluid--solid transition alone. To test this, we performed the same SLE analysis in the standard vertex model without cell division and line-tension fluctuations, and assessed SLE$_6$ consistency across three target shape indices, $p_0=3.6$ ($p_0<p^\ast$), $p_0=3.81$ ($p_0=p^\ast$), and $p_0=4.0$ ($p_0>p^\ast$), where $p^\ast \approx 3.81$ is the canonical rigidity threshold of the vertex model~\cite{bi2015density,krajnc2020solid}. As shown in \SI{SI Appendix, Fig.~S19}, SLE$_6$-like signatures emerge only for $p_0>p^\ast$, indicating that the onset of conformal behavior does not coincide exactly with the rigidity threshold. In this light, our results highlight a role for cell division that extends beyond shifting tissue fluidity: Division both reshapes the geometric and mechanical landscape of the tissue to facilitate rearrangements and, as an intrinsic turnover event, injects dynamic renewal into the cell layer, thereby sustaining the universal flow organization.

Beyond this general mechanism observed across Mitomycin C treatment, Nocodazole treatment, and vertex model simulations, we also identified the emergence of anisotropic correlations specifically in Mitomycin C-treated monolayers. 
This suggests that, in some cases, cell division not only maintains flow rearrangements but also contributes to the isotropic organization of flow structures required for the rotational symmetry encompassed by conformal invariance. 
However, this anisotropic alteration is not universal, as it is absent in Nocodazole-treated tissues.
This distinction likely reflects differences in the biological mechanisms of action: Mitomycin C interferes with DNA synthesis and causes cell cycle arrest, while Nocodazole disrupts microtubule polymerization \cite{beswick2006nocodazole}, preventing mitotic spindle formation. 
Thus, our results point to two complementary routes by which conformal symmetry and universality can be destabilized in living matter: 
a general route in which cell division alters the geometric and mechanical organizations of the cell layer, and a secondary route through the emergence of flow anisotropy that breaks rotational invariance. Importantly, applying the same anisotropic-correlation analysis to our vertex-model simulations reveals no detectable anisotropy (\SI{SI Appendix, Fig.~S20}). This confirms that the mechanics--geometry--rearrangement--conformal chain revealed by the model does not rely on anisotropy, and instead reflects the common, treatment-independent effect shared by both proliferation-inhibition perturbations.

Biologically, our findings underscore an important role for cell division: \rev{beyond driving proliferation, tissue expansion, and fluidization, it also contributes to the structured, multiscale organization of collective tissue dynamics. 
In the mechanobiological context, collective cellular motion, including the vortical motion we analyze here, is tightly coupled to mechanochemical transduction \cite{ladoux2017mechanobiology}.
Through this coupling, the conformally invariant flow we identify can serve as a mechanistic basis for the coherent organization of local and long-range mechanochemical patterns that shape tissue biology \cite{bailles2022mechanochemical}.
From a systems-biology standpoint, living systems are widely thought to favor critical regimes \cite{myrov2026hierarchical,nykter2008gene,de2017critical} because such regimes balance robustness and adaptability: They keep statistical properties stable under variations in conditions and dynamics while allowing the system to adapt to novel situations and maintain homeostasis \cite{mora2011biological,munoz2018colloquium}.
In this context, the emergence of universal, conformally invariant behavior provides a concrete manifestation of such robustness in cell layers, by showing that cell division provides a flow backbone for collective organization that follows critical percolation statistics, and our results provide direct evidence that cell division plays a key role in maintaining it.} 

These findings also carry important practical implications, as antiproliferative interventions \cite{szabo2016synthesis,chen2021significant}, commonly employed in cancer and cardiovascular studies and therapies, can exert unanticipated effects on intrinsic collective behaviors and tissue mechanics. Finally, by placing tissue dynamics in a well-defined universality class, our results open the door to using the theoretical and quantitative tools of two-dimensional critical phenomena to analyze the geometry, mechanics, and dynamics of living systems. 

\matmethods{Confluent Madin--Darby Canine Kidney (MDCK) epithelial monolayers were imaged by time-lapse phase-contrast microscopy at 10$\times$ magnification under control conditions, after cell division inhibition with Mitomycin C (Sigma-Aldrich, M4287) or Nocodazole (Sigma-Aldrich, SML1665), and, where indicated, after apoptosis inhibition with Q-VD-OPh (MedChemExpress, HY-12305). Velocity fields were extracted from the image sequences using particle image velocimetry (PIV)~\cite{thielickePIVlabUserfriendlyAffordable2014} and were then used to compute the vorticity field. Candidate traces were extracted for SLE diagnostics from zero-vorticity interfaces identified in binarized vorticity fields~\cite{andersen2025evidence}. These traces were analyzed using left-passage probability~\cite{schramm_percolation_2001} and the variance of the Loewner driving function~\cite{kennedy2009numerical} to assess compatibility with SLE$_6$ and estimate an effective diffusivity $\kappa$. Cell densities were quantified by segmenting cells with the Cellpose~\cite{stringer2021cellpose}, and Tissue Analyzer~\cite{aigouy2016segmentation} was used to quantify division and extrusion rates. Cell-center adjacency graphs were constructed using an $\alpha$-shape complex~\cite{Edelsbrunner1994Three-dimensionalShapes,Edelsbrunner1995,edelsbrunner_alpha_2011} based on cell centroids and neighbor distances, from which we quantified the mean adjacent neighbor number and the fraction of cells belonging to the largest adjacent component~\cite{jacobs1995generic,jacobs1997algorithm,zhang2015rigidity,petridou2021rigidity}. Vertex-model simulations with cell division were adapted from refs.~\citenum{bocanegra2023cell} and~\citenum{guerrero2019neuronal}. Detailed experimental procedures, SLE analysis, image analysis, simulations, and statistical methods are provided in \SI{SI Appendix}.}

\showmatmethods{} % Display the Materials and Methods section
\section*{Data, Materials, and Software Availability.}
The raw experimental datasets for control, Mitomycin C–treated, Nocodazole-treated, and Q-VD-OPh–treated cell layers, as well as the raw vertex model simulation datasets that support the findings of this study, are available at \url{https://sid.erda.dk/sharelink/DYCtyu311c}~\cite{ma2025rawdata}. The SLE analyses code is available at \url{https://gitlab.nbi.ku.dk/active-intelligent-matter/sle_division}~\cite{ma2026slecode}. All other data are included in the manuscript and/or supporting information.

\acknow{It is a pleasure to acknowledge helpful conversations with Mehran Kardar, Manu Prakash, Erwin Frey, and Lakshmi Balasubramaniam. M.C.P. thanks Villum Fonden (Grant No. 69081) for their support. N.A.M.A. acknowledges funding from the Portuguese Foundation for Science and Technology (FCT) under contract nos. EXPL/FIS-MAC/0406/2021, UIDB/00618/2020, and UIDP/00618/2020. A.D. acknowledges funding from the Novo Nordisk Foundation (Grant No. NNF21OC0068687), Villum Fonden (Grant no. 29476), and the European Union (ERC, PhysCoMeT, 101041418). Views and opinions expressed are however those of the authors only and do not necessarily reflect those of the European Union or the European Research Council. Neither the European Union nor the granting authority can be held responsible for them. We used ChatGPT (OpenAI) to help improve the clarity and readability of part of the manuscript after the initial draft was completed.}

\showacknow{}

\bibsplit[24]

% Published reference text, fixed to the supplied PNAS PDF. No BibTeX run is required.

\end{document}